\documentclass[twocolumn,astrosymb]{aastex631}

\newcommand{\oiline}{$\mathrm{[OI]}\,6300\,\mathrm{\angstrom}\,$}
\newcommand{\Neline}{$\mathrm{[NeII]}\,12.8\,\mathrm{\micron}\,$}
\usepackage{graphicx}	
\usepackage{amsmath}	
\usepackage{multirow}
\usepackage{layouts}

\graphicspath{{./}{figures/}}

\newcommand{\angstrom}{\mbox{\normalfont\AA}}

\defcitealias{Thanathibodee2023}{T23}

\newcommand{\prodimo}{P{\footnotesize RO}D{\footnotesize I}M{\footnotesize O}\;}

\received{August 11, 2023}
\revised{August 28, 2023}
\accepted{August 31, 2023}
\submitjournal{ApJL}

\DeclareRobustCommand{\VAN}[3]{#2}
\let\VANthebibliography\thebibliography
\def\thebibliography{\DeclareRobustCommand{\VAN}[3]{##3}\VANthebibliography}

\begin{document}


\title[Resolved diagnostics in TW Hya]{High-resolution [OI] line spectral mapping of TW Hya consistent with X-ray driven photoevaporation.}

\author[0000-0003-1817-6576]{Christian Rab}
\affiliation{University Observatory, Faculty of Physics, Ludwig-Maximilians-Universit\"at M\"unchen, Scheinerstr. 1, 81679 Munich, Germany}
\affiliation{Max-Planck-Institut f\"ur extraterrestrische Physik, Giessenbachstrasse 1, 85748 Garching, Germany}

\author[0000-0002-4983-0422]{Michael Weber}
\affiliation{University Observatory, Faculty of Physics, Ludwig-Maximilians-Universit\"at M\"unchen, Scheinerstr. 1, 81679 Munich, Germany}
\affiliation{Exzellenzcluster `Origins', Boltzmannstr. 2, 85748 Garching, Germany}

\author[0000-0003-3754-1639]{Giovanni Picogna}
\affiliation{University Observatory, Faculty of Physics, Ludwig-Maximilians-Universit\"at M\"unchen, Scheinerstr. 1, 81679 Munich, Germany}

\author{Barbara Ercolano}
\affiliation{University Observatory, Faculty of Physics, Ludwig-Maximilians-Universit\"at M\"unchen, Scheinerstr. 1, 81679 Munich, Germany}
\affiliation{Exzellenzcluster `Origins', Boltzmannstr. 2, 85748 Garching, Germany}

\author[0000-0002-4856-7837]{James Owen}
\affiliation{Astrophysics Group, Imperial College London, Blackett Laboratory, Prince Consort Road, London SW7 2AZ, UK}

\begin{abstract}
Theoretical models indicate that photoevaporative and magnetothermal winds play a crucial role in the evolution and dispersal of protoplanetary disks and affect the formation of planetary systems. However, it is still unclear what wind-driving mechanism is dominant or if both are at work, perhaps at different stages of disk evolution. Recent spatially resolved observations by \citet{Fang2023a} of the \oiline spectral line, a common disk wind tracer, in TW Hya revealed that about 80\% of the emission is confined to the inner few au of the disk. 
In this work, we show that state-of-the-art X-ray driven photoevaporation models can reproduce the compact emission and the line profile of the \oiline line. Furthermore, we show that the models also simultaneously reproduce the observed line luminosities and detailed spectral profiles of both the \oiline and the \Neline lines. While MHD wind models can also reproduce the compact radial emission of the \oiline line they fail to match the observed spectral profile of the \oiline line and underestimate the luminosity of the \Neline line by a factor of three. 
We conclude that, while we cannot exclude the presence of an MHD wind component, the bulk of the wind structure of TW Hya is predominantly shaped by a photoevaporative flow. 
\end{abstract}
\keywords{}



\section{Introduction}\label{sec:intro}
The evolution and final dispersal of protoplanetary disks is thought to strongly affect the formation and evolution of planetary systems. Disk winds are considered to be significant contributors to the evolutionary processes occurring within protoplanetary disks \citep{Lesur2022, Pascucci2022}. Thermal winds can be launched through photoevaporation (PE) from the central star \citep[e.g.][]{Gorti2009, Nakatani2018, Picogna2021, Ercolano2021} and are efficient at removing material at rates comparable to the observed accretion rates of T-Tauri stars \citep[e.g.][]{Ercolano2017}. Thermal winds do not remove angular momentum from the disk, and, when combined with viscous accretion models, they are successful in reproducing the observed two-timescale behavior, evidenced by the evolution of disk colors \citep[e.g.][]{Koepferl2013,Ercolano2015b} and several observational correlations such as the observed accretion rates and the mass of the central star \citep{Ercolano2014} or inner disk life times \citep{Picogna2021}. 

The inclusion of non-ideal magneto-hydrodynamical effects in weakly ionized protoplanetary disk has shown that magnetorotational instability (MRI) \citep{Balbus1991}, hypothesized to drive viscosity in disks, is suppressed in most regions of the disk \citep[see e.g.][for a recent review]{Lesur2022}, except the very inner regions where thermionic emission from dust dominates \citep{Desch2015, Jankovic2021}. Vigorous, magnetically supported, disk winds (from now on MHD winds), are a solid prediction of most recent simulations \citep[e.g.][]{Gressel2015,Bai2016,Wang2019,Lesur2021,Gressel2020} and they replace MRI in most disk regions by removing angular momentum from the disk, allowing material to flow inward.

Which type of wind might dominate at different times and different locations in a disk is an important question, which directly affects planet formation models. The current picture emerging from the careful analysis of spectroscopic diagnostics is that both types of winds operate in disks, with MHD winds stronger in young objects and thermal winds dominating the final evolution and eventual dispersal of disks \citep{Ercolano2017, Weber2020}. 

Currently used spectroscopic wind diagnostics, particularly the \oiline collisionally excited spectral line, have complex line profiles \citep[e.g.][]{Simon2016,Fang2018,Banzatti2019,Gangi2020}, often preventing important wind parameters, like the wind launching radius, to be directly determined (see discussion in \citealt{Weber2020}). \citet{Rab2022} find that a combination of \oiline and molecular hydrogen observations are consistent with thermal winds driven by X-ray photoevaporation, but alternative models cannot be ruled out. 

In order to break the degeneracies hidden in the interpretation of non-spatially resolved line profiles, high-resolution spectral mapping of wind diagnostics represent an attractive option. This has recently been done for the \oiline line from TW Hya by \citet{Fang2023a}, using the multi-unit spectroscopic explorer (MUSE) at the Very Large Telescope, who showed that about 80\% of the [OI] emission is confined to within 1~au radially from the star. In this paper we show that state-of-the-art thermal wind models driven by X-ray photoevaporation \citep[e.g.][]{Picogna2019,Ercolano2021,Picogna2021} are consistent with the observations recently published by \citet{Fang2023a}. In Section \ref{sec:methods} we briefly describe the used photoevaporative disk wind models and our approach to produce synthetic observables. In Section \ref{sec:results} we show our results in particular the comparison to the observational data. We discuss our results in context to previous works and MHD disk wind models and present our conclusions in Section \ref{sec:conclusions}. 
\section{Methods}\label{sec:methods}
In this section, we describe the physical models used in this work and how we produce synthetic observables from those models that can be directly compared to observational data. 
\subsection{Photoevaporative disks wind models}
To model a photoevaporative disk wind we follow the approach by \citet{Picogna2019, Picogna2021, Ercolano2021}\footnote{X-ray PE models and data from \citet{Picogna2021} are available here \url{https://cutt.ly/lElY9JI}}, to which we refer for details. This model uses a modified version of the PLUTO code \citep{Mignone2007, Picogna2019} to perform radiative-hydrodynamic simulations of a disk irradiated by a central star. The temperatures in the wind and the wind-launching regions, i.e. the upper layers of the disk, where the column number density towards the central star is in the range between $5\times 10^{20}$ and $2.5\times 10^{22}$ cm$^{-2}$, are determined by parameterizations that are derived from detailed radiative transfer calculations with the gas photo-ionization code MOCASSIN \citep{Ercolano2003a, Ercolano2005, Ercolano2008a}. For a given column number density the respective parameterization yields the gas temperature dependent on the ionization parameter $\xi = \frac{L_X}{n r^2}$, where $L_\mathrm{X}$ is the X-ray luminosity of the star, $n$ the volume number density and $r$ the spherical radius. 
In this work, we use a stellar mass $M_* = 0.7 M_\odot$ and the parametrizations derived from the spectrum labeled as Spec29 in \citet{Ercolano2021} with $L_X = 2\times 10^{30}$ erg s$^{-1}$, which is appropriate considering observational constraints on TW Hya \citep{Robrade2006,Ercolano2017b}. The computational grid was centered on the star. Spherical polar coordinates were adopted with 512 logarithmic spaced cells in the radial direction from $0.33$ au to $600$ au, and 512 uniform spaced cells in the polar one from $0.005$ to $\pi/2$. Outflow boundaries were adopted in the radial directions, while special reflective boundaries were used to treat the regions close to the polar axis and the disk mid-plane. A periodic boundary was assumed in the azimuthal direction. The influence of the inner boundary was tested by decreasing the inner radial boundary to $0.1$ au, while keeping the same radial resolution outside $0.33$ au.

\subsection{Disk model without a wind}
Additionally to the PE disk wind models, we use an existing radiation thermo-chemical disk model for TW Hya from the DIANA (DIsc ANAlysis\footnote{\url{https://diana.iwf.oeaw.ac.at}}) project presented in \citet{Woitke2019}. This model does not include a wind component but was made to reproduce existing (mostly spatially unresolved) observational data including the spectral energy distribution and about 50 spectral lines (i.e. line fluxes are matched within a factor of two to three). With this model we show how a pure disk model compares with the spatially resolved \oiline observations. As at the time of the publication of this model no spatially resolved observables for the \oiline were produced, we rerun the model with a more recent version of the radiation thermo-chemical code \prodimo (PROtoplanetary DIsk MOdel\footnote{\url{https://prodimo.iwf.oeaw.ac.at} \mbox{revision: 66efbd75 2023/06/27}}, \citealt{Woitke2009a,Kamp2010,Thi2011,Woitke2016}) to produce line cubes and images that can be compared to the spatially resolved data.
\subsection{Synthetic observables}
To produce synthetic observables we use two different approaches to post-process the PE disk wind model. Similar to \citet{Weber2020} we use the MOCASSIN Monte Carlo radiative transfer code that allows to model spectral line emission for atomic species in the optical and infrared \citep{Ercolano2003a,Ercolano2005,Ercolano2008a}. Furthermore we use the radiation thermo-chemical code \prodimo that was recently applied in \citet{Rab2022} to produce synthetic observables for atomic and molecular species that are supposed to trace disk winds. We use both approaches to show that our results, in particular the spatial extent of the \oiline, are robust and do not strongly depend on the details of the post-processing method (e.g. chemistry, heating-cooling, line excitation). 

For both approaches, we use the same physical structure (density and velocity field), the same X-ray/stellar spectrum, and the same dust properties. To account for the accretion luminosity we add to the X-ray spectrum a black body spectrum with $T_\mathrm{eff}=12000\,\mathrm{K}$ normalized to \mbox{$L_\mathrm{acc}=2.95 \times 10^{-2}~L_\odot$} \citep{Fang2018}. For the dust we assume a gas-to-dust ratio of 100 and an ISM size distribution (see \citealt{Weber2020} for details). 
Quantities such as temperatures, line populations, chemical abundances and the synthetic observables are self-consistently calculated within each post-processing framework. For a more detailed discussion on these different post-processing approaches and a comparison see \citet{Rab2022}.

At first we use the line radiative transfer modules of MOCASSIN and \prodimo to produce synthetic model images for the \oiline emission assuming a distance of $60\,\mathrm{pc}$ \citep{Gaia2021} and a disk inclination of $7^\circ$ \citep{Qi2004}, consistent with \citet{Fang2023a}. For the produced model images we use an oversampling factor of seven ($0.003615\arcsec$) compared to the pixel scale of the observations. Following \citet{Fang2023a} we downsample the model images to the pixel scale of the MUSE data before convolving it with the PSF. For these model images and for the observed image we produce azimuthally averaged radial profiles using the corresponding pixel scale as the width for the radial bins. Additionally we also produce radial profiles for the unconvolved model images to better show the real extend of the emission. We also produce synthetic images in the same way for the toy model (power-law model) presented in \citet{Fang2023a} and the thermo-chemical model without a wind. We note that the used oversampling factor is not enough to fully resolve the toy model of \citet{Fang2023a} but we found that it is sufficient to reproduce the observation after convolution of the toy model image (mainly because a downsampling to the pixel size of the observations is anyway required). We also note that a higher resolution spatial grid is used for the radiative transfer step, but as we are not aiming for a detailed fitting of the observational data we use the oversampling factor of all model images for consistency (e.g. between the two different post-processing methods) and efficiency. 

We also compare our model with the observations of the \Neline line presented in \citet{Pascucci2011}. For this line we simply produce spectral line profiles again using both post-processing approaches. 

\section{Results}\label{sec:results}
Here, we compare our modeling results to the observational data. As we do not present newly developed PE models we focus only on the observables. In Sect.~\ref{sec:radprofs} we present normalized azimuthally averaged radial profiles for \oiline in a similar fashion as \citet{Fang2023a} and in Sect.~\ref{sec:specprofs} we compare our PE wind models to the observed spectral profiles for the \oiline and \Neline spectral lines. 

\subsection{Radial profiles}
\label{sec:radprofs}
\begin{figure}
    \centering
    \includegraphics[width=0.48\textwidth]{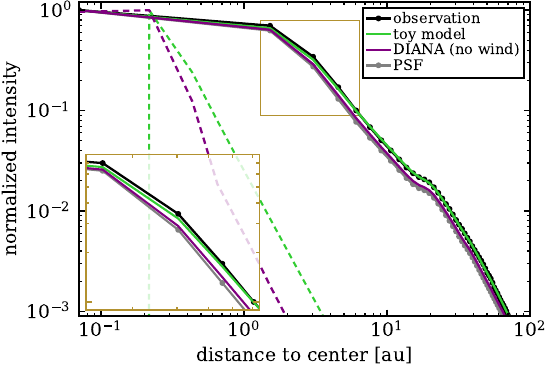}
    \caption{The toy (power-law) model (green) from \citet{Fang2023a} compared to results from a radiation thermo-chemical disk model (purple) without a wind that fits a large set of observational data for \mbox{TW Hya} \citep{Woitke2019}. The dashed and solid colored lines show the radial profiles for the unconvolved and convolved images, respectively. The gray solid line shows the radial profile of the PSF and the black solid line for the \oiline MUSE observations \citep{Fang2023a}. The inset shows a zoom-in to the region marked by the brown box ($r=1.3-6.4\,\mathrm{au}$) in the main panel.}
    \label{fig:toymodelcomp}
\end{figure}
\subsubsection{Disk only model}
In Fig.~\ref{fig:toymodelcomp}, we show the toy model from \citet{Fang2023a} in comparison to the disk-only model from \citet{Woitke2019}. The thermo-chemical disk model shows a similar radial profile for the \oiline emission, in particular the steep slope, but is slightly more compact and hence does not match the data (i.e. it remains unresolved). We note that the model of \citet{Woitke2019} uses a parameterized disk structure, which is quite different to the disk wind models used in this work or in \citet{Fang2023a}, in particular, it has a dust and gas depleted (optically thin but not empty) inner hole extending up to almost $3\,\mathrm{au}$. Although it is possible to adapt this model to achieve a better match to the spatially resolved data, such a disk only model, by construction, cannot match the blue-shifts seen in the observed line profiles of the \oiline and \Neline lines. Nevertheless, it is interesting to see that such a model is almost in agreement with the spatially resolved data, although such constraints were not included in the modeling. Furthermore, this model indicates that a disk only solution seems to produce an even more compact emission region for the \oiline compared to the wind models presented here or in \citet{Fang2023a} and also that the contribution from a disk itself can be significant in the case of TW Hya.
\begin{figure}
    \centering
    \includegraphics[width=0.48\textwidth]{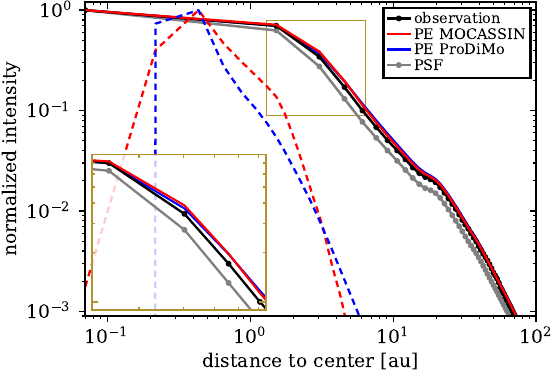}
    \caption{Comparison of the photoevaporative disk wind model with the observed normalized radial intensity profiles of the \oiline line. The black solid line shows the observation, the gray solid line the PSF (both from \citealt{Fang2023a}. The colored solid and dashed lines show the model results, convolved with the PSF and at the resolution of the model image, respectively. In red, we show the results from the MOCASSIN and in blue from the \prodimo post-processing. The inset shows a zoom-in to the region marked by the brown box ($r=1.3-6.4\,\mathrm{au}$) in the main panel.}
    \label{fig:mocmodelcomp}
\end{figure}
\begin{figure}
    \centering
    \includegraphics[width=0.47\textwidth]{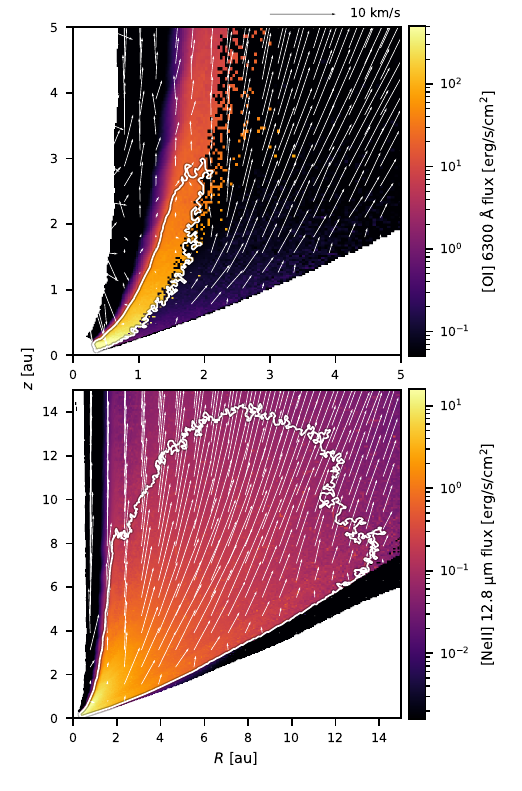}
    \caption{\oiline (top) and \Neline (bottom) line flux of our photoevaporation disk wind model obtained by multiplying the emissivity with $2 \pi R$ in order to illustrate the origin of the emission. White solid contour lines indicate the regions where the top 80 \% of emission originates. White arrows represent the velocity field in the PE model.}
    \label{fig:emis-regions}
\end{figure}
\subsubsection{Photoevaporative disk wind models}
In Fig.~\ref{fig:mocmodelcomp} we compare the \oiline radial intensity profiles of our PE wind model at an inclination of $7^{\circ}$ to the observation by \citet{Fang2023a}. As can be seen in the unconvolved radial profiles, both, the MOCASSIN and \prodimo models yield very similar results, with the \prodimo model having slightly more emission at very low ($\lesssim 0.4$~au) and at extended radii, and MOCASSIN showing enhanced emission at intermediate radii between $\approx 0.5$ and 3~au. In both models, the emission peaks well inside of 1 au with a steep decrease in intensity at larger radii. Computing the cumulative integral, we find that 80\% of the emission originates inside 2~au of the central star, compared to 1~au for the toy model that \citet{Fang2023a} derived as a fit to the data. This is also visible in Fig.~\ref{fig:emis-regions}, where we show 2D emission maps overlain by contours showing the 80\% regions. It is worth pointing out that although the 80\% regions extend to $\approx 2$~au and $\approx 14$~au for the \oiline and \Neline lines, respectively, the emission inside this region is not uniform but has a strong gradient with the peak close to the star.
Comparing the profiles after convolution with the PSF, it can be seen that the radial profiles of the PE wind model are in very good agreement with the observations. This shows that current state-of-the-art photoevaporative disk models produce compact emission consistent with the spatially resolved observations of \mbox{TW Hya}.

Figure \ref{fig:emis-regions} shows significant emission originating from the inner 1 au, which is inside the gravitational radius of the X-ray photoevaporation models shown here ($\sim 3.5$ au). The emitting material close to the star is thus bound to the inner disk and not affected by photoevaporation. This implies that our main result would not qualitatively change in the case of a "gapped" inner disk as suggested by the observations of a dark annulus in mm-wave dust emission at \mbox{$\approx$ 1 au} \citep{Andrews2016}. Indeed, the presence of gas close to the central star of TW Hya, which shows clear accretion signatures,  justifies the employment of primordial disk models in this case. While we do not expect significant changes to the wind structure due to the gap and hence on the main picture of having compact emission, a detailed hydrodynamical model would be required to assess potential differences in the detailed spectroscopic line profiles emitted from this region (see also Sect.~\ref{sec:conclusions}). Such a model is, however, outside the scope of this study and would also require higher spatial resolution in the observations of the inner disc, which is not possible with current instrumentation.
\begin{figure*}
    \centering
    \includegraphics{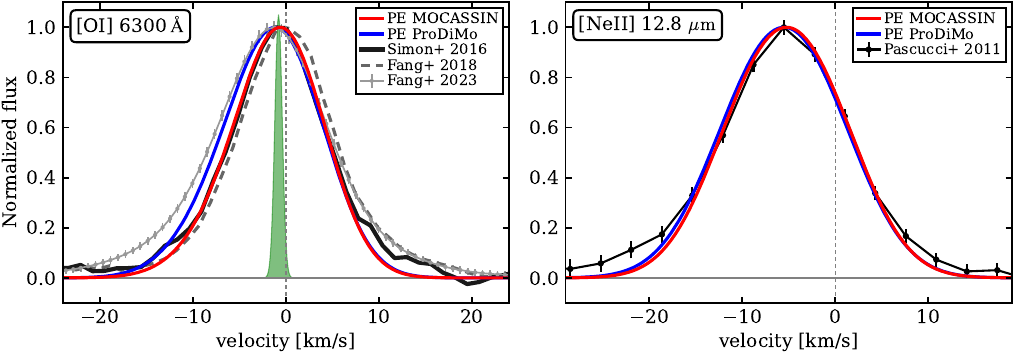}
    \caption{Comparison of observed \oiline \citep{Simon2016,Fang2018,Fang2023a} and \Neline \citep{Pascucci2011} spectral line profiles to synthetic line profiles of photoevaporative disk wind models. The [OI] line profiles from the models and that observed by \citet{Fang2023a} have been degraded to a resolving power $R = 40000$, comparable to the resolution in the observation by \citet{Simon2016}. For the \citet{Fang2023a} data the error bars correspond to $3\sigma$ uncertainties (for visibility). The Gaussian distribution shown in green represents the derived mean centroid shift of $-0.8\,\mathrm{km s^{-1}}$ with a standard deviation of $0.4\,\mathrm{kms^{-1}}$ as derived by \cite{Fang2023a} from various observations of \oiline. The [NeII] model profile has been degraded to $R=30000$ and the error bars in the observed profile indicate $1\sigma$ uncertainties.}
    \label{fig:profile_oi}
\end{figure*}
\subsection{Spectral line profiles}
\label{sec:specprofs}
In Fig.~\ref{fig:profile_oi} we compare the modelled spectral line profiles to the observations of the \oiline and \Neline spectral lines. For \oiline we show three different observed spectra representing the scatter observed in the measured centroid velocities $v_\mathrm{c}$ with a mean of $v_\mathrm{c}=-0.8\,\mathrm{km s^{-1}}$ and a standard deviation of $0.4\,\mathrm{km s^{-1}}$ \citep{Fang2023a}. Both models are in good agreement with the observations. In particular the MOCASSIN ($v_c\approx-0.8\,\mathrm{km s^{-1}}$) model matches the profile of \citep{Simon2016}, that shows $v_c\approx-0.8\,\mathrm{km s^{-1}}$, exceptionally well. The \prodimo profile is slightly broader in the blue part of the spectrum and therefore appears more blue-shifted ($v_c\approx-1.5\,\mathrm{km s^{-1}}$). Nevertheless this is still consistent with observations as for example the profile of \citet{Fang2023a} shows a similar behavior with $v_c\approx-1.5\,\mathrm{km s^{-1}}$, for the spectrum downgraded to R=40000. The \oiline line fluxes from the models are $1.2\times10^{-5}\,L_\sun$ (MOCASSIN) and $5.5\times10^{-6}\,L_\sun$ (\prodimo), in good agreement with the observed values of $1.0-1.5\times10^{-5}\,L_\sun$ \citep{Simon2016,Fang2018,Fang2023a}.

The difference in the shape of the two model spectra can be explained by the slightly more extended \oiline emission of the \prodimo model (see Fig.\ref{fig:mocmodelcomp}), which traces slightly faster velocities of the PE wind compared to the MOCASSIN model. We tested this by simply removing all emission for $r>4\,\mathrm{au}$ in the synthetic observables of the \prodimo model and find that for this case the MOCASSIN and \prodimo line profiles become almost identical. As the density structure and velocity fields in both models are the same; the differences arise from different radial temperature gradients and differences in the line excitation calculations. However, our results indicate that detailed models for TW Hya, fully considering the stellar properties and possibly also details of the disk structure (i.e. a gap in the disk structure; see \citealt{OwenPhD2011}) are required for a comprehensive interpretation of the \oiline line profile. 

For \Neline the modeled profiles are very similar and match very well the observed spectral profile. As \Neline traces regions further out and higher up (up to r $\approx10\,\mathrm{au})$ in the disk/wind with respect to \oiline, it traces higher velocities of the photoevaporative flow (see Fig.~\ref{fig:emis-regions}), consistent with the observed $v_\mathrm{c}\approx -5 \,\mathrm{km s^{-1}}$. The line luminosity of the models is $3.4\times10^{-6}\,L_\sun$ (MOCASSIN) and $5.1\times10^{-6}\,L_\sun$ (\prodimo) which are in excellent agreement with the observed range of luminosities of $\approx\!3.5-6.2\times10^{-6}\,L_\sun$ \citep{Pascucci2011,Pascucci2009,Najita2010}. 

\section{Discussion and Conclusions}\label{sec:conclusions}
Our results show that current state-of-the-art PE disk wind models are consistent with the observational data presented in \citet{Fang2023a}. We want to emphasize that for this work, we did not use any new developments or adaptations to the PE wind modeling approach already presented and used in \citet{Picogna2019,Weber2020,Picogna2021,Ercolano2021,Rab2022,Weber2022}. The only difference between this work and previously published work is the choice of an appropriate X-ray spectrum for TW Hya. 

\citet{Fang2023a} compared the spatially resolved \oiline emission to very low resolution early photoevaporation models \citep{Owen2010,Ercolano2010} which present a more extended emission profile for the \oiline. Based on this comparison, they concluded that a magneto-thermal driven wind is necessary to explain the spatially resolved \oiline line data. We show here, however, that, compared to the fiducial magneto-thermal wind model presented in \citet{Fang2023a}, modern PE models produce only very slightly more extend \oiline emission but are still fully consistent with the data. Apart from the higher inner grid resolution, an important difference between the "old" PE models \citep{Owen2010,Ercolano2010} and the newer PE models \citep{Picogna2019,Weber2020,Picogna2021,Ercolano2021} is the temperature parameterization. The new models take into account the detailed column density distribution (attenuation) in the simulation regions, yielding a more accurate temperature (and thus density) profile. More specifically, the density in the inner disk is higher for the new models, resulting in a more compact emission region.

The new PE models can match the observed line luminosity and the shape of the \oiline profile very well, in particular the small observed blue-shift of $v_c\approx-0.8\,\mathrm{km/s}$. As noted by \citet{Fang2023a} the \oiline spectral profile of their MHD model is too blue-shifted compared to observations, which is likely caused by the higher wind velocities in the inner regions compared to our PE model. \citet{Fang2023a} argue that the discrepancy could be caused by the lack of an inner hole in their primordial disk model and that the presence of such a hole would allow for red-shifted emission from the back side of the disk to contribute to the line profile, reducing its blue-shift. However, in order to zero the velocity center by this effect, the line would then result broadened by the blue-shifted value \citep{Ercolano2010, Ercolano2016}, which would then again be in tension with the observations. Additionally, as TW Hya is seen almost face-on, the contribution from the back-side of the disk will be limited unless an unsuitable large inner hole (several au) is assumed. More quantitatively speaking, our model has an inner radius of $r_\mathrm{in}=0.33\,\mathrm{au}$ (i.e. larger than the MHD model), but still the back side of the disk contributes to less than 2\% to the total flux and does not have any significant impact on the centroid velocity. We also tested a PE wind model with $r_\mathrm{in}=0.1\,\mathrm{au}$ and found no significant differences. In any case, for a more thorough interpretation of the \oiline spectral profile of TW Hya, a more detailed disk structure model for the inner few au is required. In particular constraints from ALMA observations (gap at $\approx\,1\,\mathrm{au}$ and unresolved emission from $r<0.5\,\mathrm{au}$; \citealt{Andrews2016}$)$ and VLT/SPHERE (marginal detection of emission from the inner few au; \citealt{vanBoekel2017}) indicate that there is likely still some dust in the inner $1-2\,\mathrm{au}$ \citep[see also][]{Ercolano2017b} that would reduce \oiline emission from the backside of the disk.

As already noted in \citet{Pascucci2011} and also discussed in \citet{Fang2023a} the \Neline line observation points towards a thermally driven wind. This is supported by the models presented in this work, as the agreement of the PE wind model with the spectral line profile and the observed line fluxes (which is under-predicted by a factor of three by the MHD model of \citealt{Fang2023a}) is excellent, indicating that at least for the radii larger than a few au, traced by the \Neline line, the disk wind structure of TW Hya is predominantly shaped by a PE flow. 

We conclude that, while the currently available spatially resolved data does not allow to clearly distinguish a pure thermally driven wind from a magneto-thermal wind, the simultaneous agreement of the PE wind models with the spatially resolved data and the spectral profiles of \oiline and \Neline strongly indicates that at least large parts of the disk wind seen in \mbox{TW Hya} are driven by X-ray photoevaporation. 

\vspace*{0.5cm}

{\large \textit{Acknowledgements:}} We thank the anonymous referee for a quick and constructive report. We thank Min Fang for providing their reduced observational data for the \oiline line including the PSF. We acknowledge the support of the Deutsche Forschungsgemeinschaft (DFG, German Research Foundation) Research Unit ``Transition discs'' - 325594231. This research was supported by the Excellence Cluster ORIGINS which is funded by the Deutsche Forschungsgemeinschaft (DFG, German Research Foundation) under Germany's Excellence Strategy - EXC-2094 - 390783311. CHR is grateful for support from the Max Planck Society. This research has made use of NASA's Astrophysics Data System.

\software{This research made use of Astropy, a community-developed core Python package for Astronomy \citep{AstropyCollaboration2013,AstropyCollaboration2018}. This research made use of Photutils, an Astropy package for detection and photometry of astronomical sources \citep{larry_bradley_2022_6825092}. matplotlib \href{https://doi.org/10.5281/zenodo.7637593}{version 3.7.0} \citep{Hunter2007}; scipy \citep{2020SciPy-NMeth}; numpy \citep{harris2020array}.}


\bibliography{lib}{}
\bibliographystyle{aasjournal}



\end{document}